\documentclass[pre,notitlepage,superscriptaddress,twocolumn]{revtex4-1}

\usepackage{graphicx,latexsym,amsmath,xspace,xcolor}
\usepackage{hyperref}
\usepackage{multirow}
\usepackage{textcomp}
\usepackage{mwe}
\usepackage{soul}

\def\c#1{~\cite{#1}}
\def\cc#1{Ref.\c{#1}}

\def\f#1{Fig.~\ref{#1}}

\def\s#1{Appendix~\ref{#1}}

\newcommand{\coo}{CO$_2$\xspace}

\newcommand{\et}[2]{\ensuremath{\mathsf{ #1 \to #2 }}}
\newcommand{\comm}[1]{}

\definecolor{blue}{rgb}{0,0,1}

\begin{document}

\title{Hysteresis curves reveal the microscopic origin of cooperative \coo adsorption in diamine-appended metal--organic frameworks}

\author{John R. Edison}
\email{dayakaran@brandeis.edu}
\affiliation{Molecular Foundry, Lawrence Berkeley National Laboratory, 1 Cyclotron Road, Berkeley, CA 94720, USA}
\affiliation{Martin A. Fisher School of Physics, Brandeis University, Waltham, MA 02454, USA}
\author{Rebecca L. Siegelman}
\affiliation{Department of Chemistry, University of California, Berkeley, CA 94720, USA}
\affiliation{Materials Sciences Division, Lawrence Berkeley National Laboratory, Berkeley, CA 94720, USA}
\author{Zden\v ek Preisler}
\affiliation{Molecular Foundry, Lawrence Berkeley National Laboratory, 1 Cyclotron Road, Berkeley, CA 94720, USA}
\author{Joyjit Kundu}
\email{joyjitkundu032@gmail.com}
\affiliation{Molecular Foundry, Lawrence Berkeley National Laboratory, 1 Cyclotron Road, Berkeley, CA 94720, USA}
\affiliation{Department of Chemistry, Duke University, Durham, NC 27708, USA}
\author{Jeffrey R. Long}
\affiliation{Department of Chemistry, University of California, Berkeley, CA 94720, USA}
\affiliation{Materials Sciences Division, Lawrence Berkeley National Laboratory, Berkeley, CA 94720, USA}
\affiliation{Department of Chemical and Biomolecular Engineering, University of California, Berkeley, CA 94720, USA}
\author{Stephen Whitelam}
\email{swhitelam@lbl.gov}
\affiliation{Molecular Foundry, Lawrence Berkeley National Laboratory, 1 Cyclotron Road, Berkeley, CA 94720, USA}

\keywords{metal--organic frameworks, cooperative adsorption, dynamics, hysteresis}

\begin{abstract}
Diamine-appended metal--organic frameworks (MOFs) of the form Mg$_2$(dobpdc)(diamine)$_2$ adsorb \coo in a cooperative fashion, exhibiting an abrupt change in \coo occupancy with pressure or temperature. This change is accompanied by hysteresis. While hysteresis is suggestive of a first-order phase transition, we show that hysteretic temperature-occupancy curves associated with this material are qualitatively unlike the curves seen in the presence of a phase transition; they are instead consistent with \coo chain polymerization, within one-dimensional channels in the MOF, in the absence of a phase transition. Our simulations of a microscopic model reproduce this dynamics, providing physical understanding of cooperative adsorption in this industrially important class of materials.
\end{abstract}

\maketitle

\section{Introduction} Metal--organic frameworks (MOFs) are porous, crystalline materials with large internal surface areas, and have been studied extensively as adsorbents for gas storage and separations\c{yaghi2009,long_chemrev,yaghi2013}. The recently-developed MOFs Mg$_2$(dobpdc)(diamine)$_2$ are particularly promising in this regard because they exhibit cooperative adsorption, in which a conveniently small change in pressure or temperature results in an abrupt change in the quantity of \coo adsorbed by the framework\c{mcdonald2012,david2015,long_mixture}. In a majority of cases, cooperative adsorption results from a guest-induced phase transition or dynamic rearrangement of the MOF structure, or a phase transition of the adsorbed gas\c{bon2014,emile2017,coudert2008,coudert2011,coudert2013,smit2013,simon2017statistical}. However, gas adsorption in the diamine-appended MOFs proceeds through a one-dimensional chemical reaction. As described in Ref.~\c{david2015}, proton transfer and nucleophilic attack of N on a $\rm{CO}_2$ molecule forms an ammonium carbamate species, and induces a chain reaction that leads to the cooperative insertion of $\rm{CO}_2$ into the metal-amine bonds. The result is a chain of ammonium carbamate formed along the pore axis. The one-dimensional nature of this process, and the physics of one-dimensional phenomena, suggests that cooperative gas uptake in diamine-appended Mg$_2$(dobpdc) materials occurs in the {\em absence} of a phase transition\c{binney1992theory}. Clarifying this issue is important in order to establish a microscopic understanding of cooperative gas uptake.

In previous work\c{Kundu2018} we used a statistical mechanical model of $\rm{CO}_2$ adsorption within the MOFs Mg$_2$(dobpdc)(diamine)$_2$ to show that the process of ammonium carbamate chain formation is associated with cooperative thermodynamics consistent with experimental data. That work provided evidence for the claim that a phase transition does not need to occur in this class of materials in order to observe cooperative binding. In this paper we study the {\em dynamics} of cooperative adsorption in the representative diamine-appended MOF e-2--Mg$_2$(dobpdc) (e-2 $=$ $N$-ethylethylenediamine). We show that dynamics, in the form of the hysteresis seen in temperature-occupancy curves, also indicates that cooperative adsorption occurs in the absence of a phase transition.

 Hysteresis, a memory of the prior state of the system, is indeed normally associated with a first-order phase transition, which involves a discontinuous change of an order parameter (e.g. gas occupancy) with a control parameter (e.g. temperature). In that scenario, hysteresis arises from the slow dynamics of nucleation, i.e. the potentially long time required for thermal fluctuations to produce a nucleus of the stable phase (during which time the system remains ``stuck'' in its initial state)\c{Evans1990,Monson2012}. However, hysteresis results more generally whenever a system's order parameter changes less rapidly than its control parameter, and so can also arise in the absence of a phase transition. For instance, force-extension curves for DNA stretching display hysteresis associated with the slow detachment of the two strands of the helix\c{smith1996overstretching,cocco2004overstretching,whitelam2008there,fu2010two}. Here we present experimental and simulation data showing that the nature of hysteresis in temperature-occupancy data for \coo in e-2--Mg$_2$(dobpdc) is consistent with \coo polymerization in the absence of an underlying phase transition. Our work provides a link between microscopic models of gas adsorption and experimental data, and provides fundamental understanding of a phenomenon of experimental and industrial importance.
 
\section{Results}
\subsection{Qualitative overview of hysteresis in gas-uptake data}
\begin{figure}[]
	\includegraphics[width=0.8\linewidth]{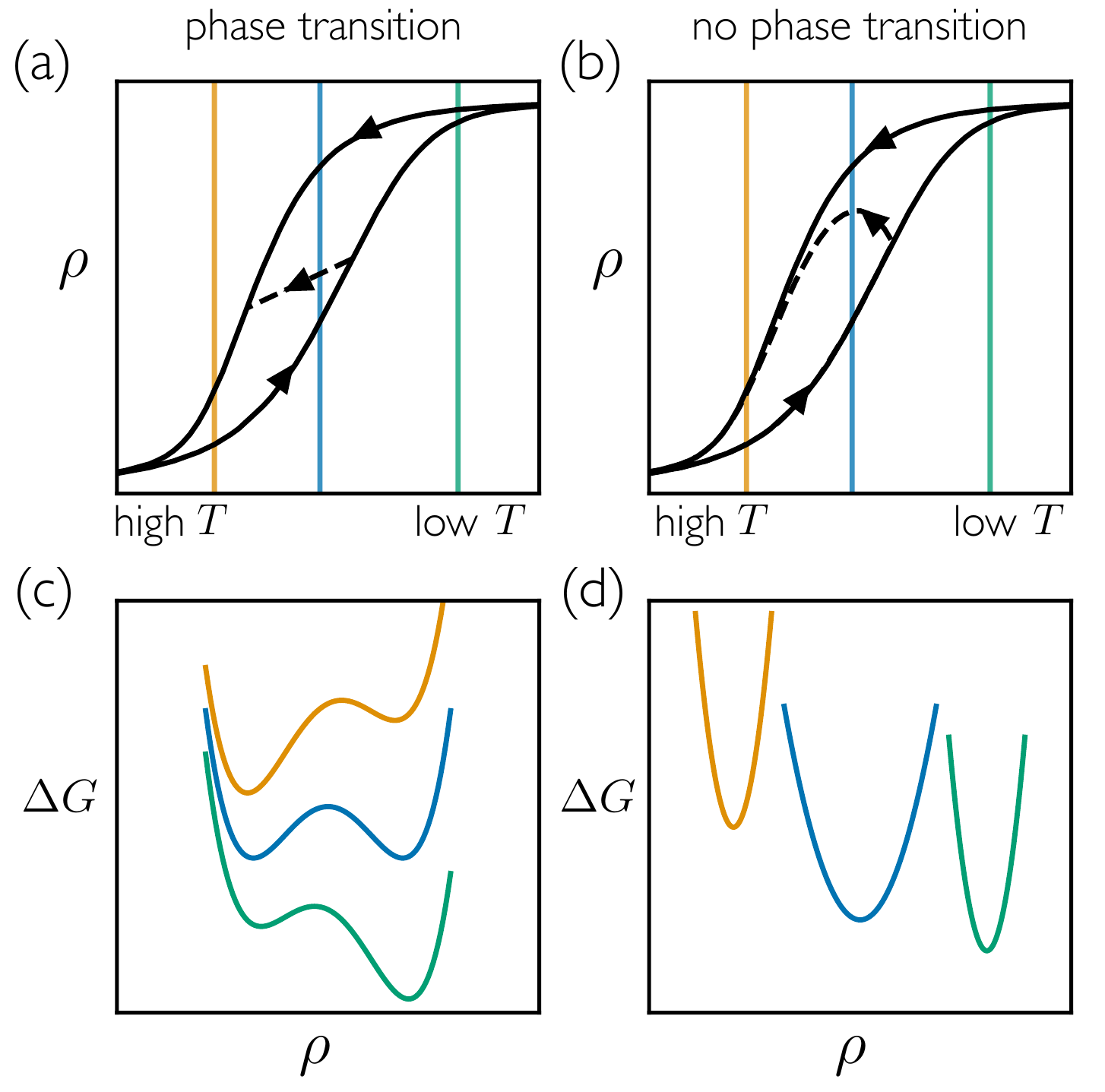}
	\caption{Schematic adsorption-desorption in the presence and absence of a phase transition. (a) Generic adsorption (up arrow) and desorption (down arrow) curves, in an occupancy $\rho$ versus temperature $T$ representation, in the presence of a first-order phase transition. Panel (c) shows the associated free-energy $(\Delta G)$ profiles. The black dashed line in panel (a) shows an ``early stop'' desorption scanning curve initiated from partway up the adsorption curve: this desorption curve decreases monotonically as we move left on the figure, because the large timescale required to access the stable state prevents the curve from moving upwards to the point corresponding to the global free-energy minimum. (b) By contrast, desorption curves in the absence of a first-order phase transition are non-monotonic, because the system can evolve toward the global free-energy minimum on the timescale of observation; panel (d) shows associated free-energy profiles. The colored vertical lines in panels (a) and (b) correspond to the similarly-colored profiles in panels (c) and (d), respectively.}
	\label{fig1}
\end{figure}

To provide a basis for our claim we summarize in \f{fig1}(a,c) the canonical case of hysteresis accompanying a first-order phase transition. In panel (a) we show typical adsorption-desorption curves~\footnote{Such curves are often referred to as ``isobars'', even though the system is not in equilibrium. The true equilibrium isobar has a unique value as a function of temperature.} for an ordered porous material with a narrow pore-size distribution, as a function of temperature $T$, for the case in which the adsorbate or framework undergoes a first-order phase transition (or a ``rounded'' transition if the system is of finite size\c{wilms2010rounding}). In panel (c) we show the accompanying free-energy landscape for a single pore\c{Evans1990,Monson2012}. 

The origin of hysteresis in this scenario is the slow dynamics of nucleation: we must wait for a thermal fluctuation to generate a nucleus of the stable phase. Near phase coexistence the nucleation time is large, and can therefore exceed the experimental observation time. However, the system retains the ability to relax to the local or metastable equilibrium as $T$ is varied\c{Debenedetti1996}. This separation of timescales can be identified by ``early stop'' desorption scans, initiated partway up the adsorption curve. One such example is shown as a black dashed line in panel (a). As $T$ increases (moving left on the figure) the adsorbate loading (dashed line) decreases monotonically: pores that are empty when the desorption scan begins remain so as the scan proceeds\c{Everett1952,Everett1954,Kruk2000,McNall2001,Tompsett2005} (see e.g. Fig. 17 of \cc{Monson2012}). Materials that display adsorption hysteresis due to an underlying structural transition show similar behavior\c{ghysels2013}.

By contrast, we expect early-stop desorption experiments in the absence of a phase transition to behave as shown in \f{fig1}(b). The associated free-energy surface, shown in panel (d), has a single minimum under all conditions, and the system should be able to evolve in the direction of this minimum on the timescale of the experiment. The result is a non-monotonic early-stop desorption curve.

\subsection{Experimental data are consistent with the absence of a phase transition}
\begin{figure*}
  	\includegraphics[width=0.8\linewidth]{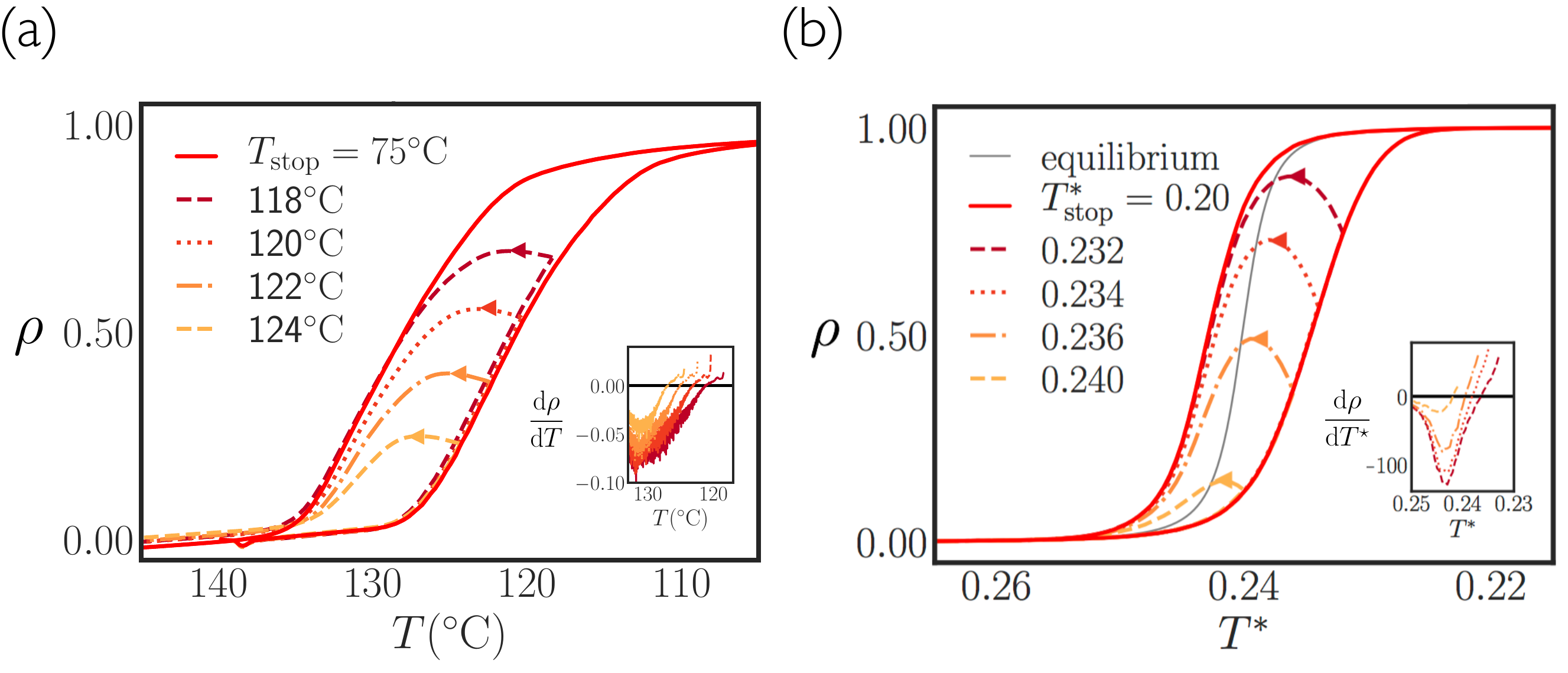}
  \caption{Early-stop desorption scanning curves for (a) \coo occupancy $\rho$ in the MOF e-2--Mg$_2$(dobpdc) and (b) our simulation model of the same are consistent with \f{fig1}(b): the qualitative nature of hysteresis indicates the {\em absence} of a phase transition in this MOF. $T_{\rm stop}$ is the temperature at which the scanning curve is reversed. Arrows point in the direction of the scanning curve. The insets to both panels show the gradients of the scanning curves, which change sign. In panel (b), $T^\star \equiv  k_{\rm B}T/\epsilon$, where $\epsilon = 22.6$ kJ/mol is a basic unit of energy. The basic unit of simulation time is determined in \s{simulations}.}
  \label{fig2} 
\end{figure*}
\begin{figure}
  \includegraphics[width=\linewidth]{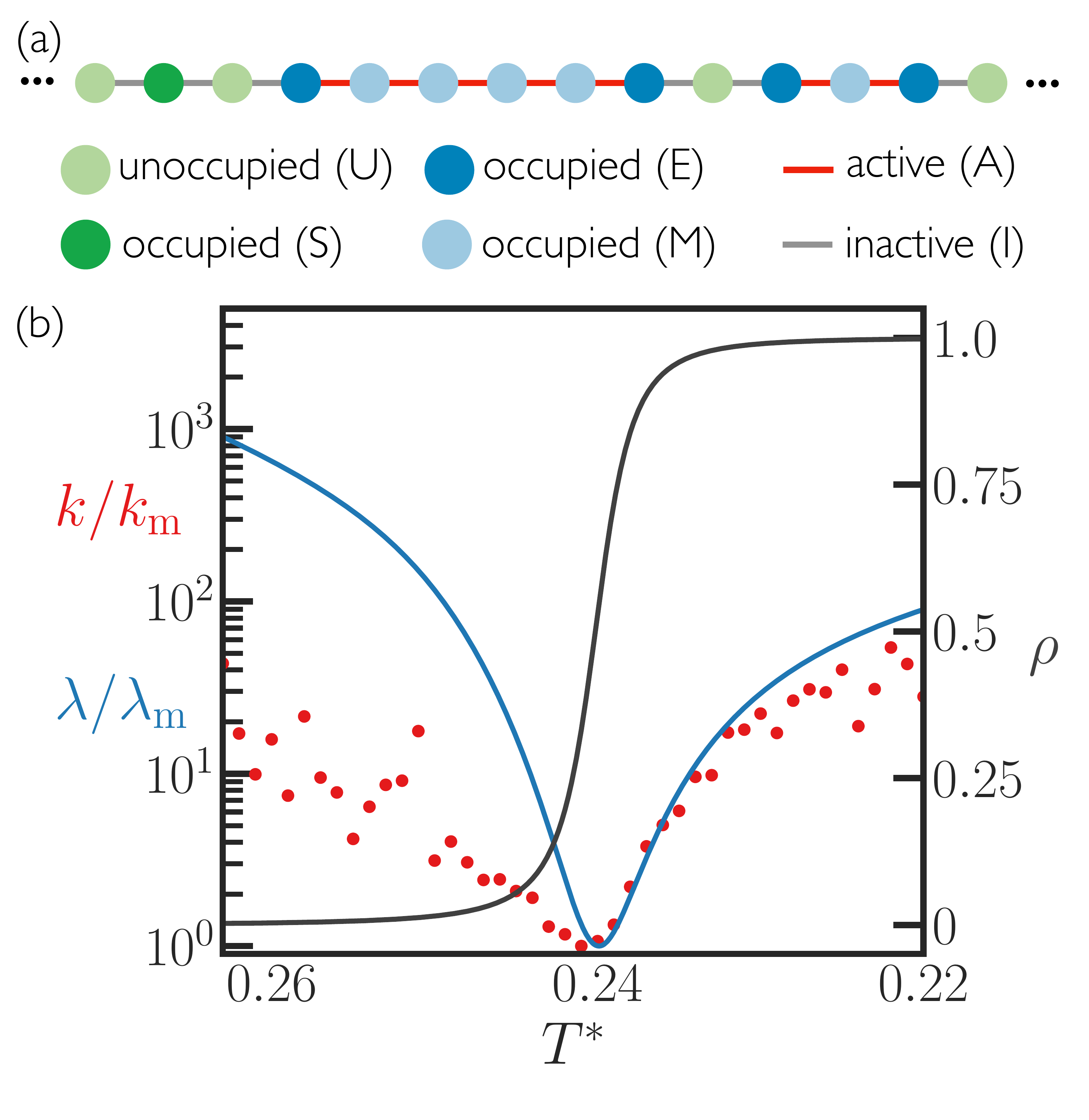}
   \caption{(a) Schematic of our statistical mechanical model of a diamine-appended MOF. Each site of the lattice can be unoccupied (U) or occupied by individual \coo molecules (S) or molecules at the end (E) or middle (M) of polymerized \coo chains. Chains are held together by active (A) bonds, while inactive (I) bonds connect all other sites. The rates for changes of these states are given in \s{simulations}. (b) The approximate curvature of the free-energy landscape of the model, $\lambda$, (blue) influences its relaxation rate $k$ (red) along the adsorption isobar (gray). At the inflection point of the isobar the flatness of the free-energy landscape results in slow dynamics and the hysteresis seen in \f{fig2}. The parameters $k_{\rm m} \approx 1.30 \times 10^{-9}$ s$^{-1}$ and $\lambda_{\rm m} \approx 0.267$ are reference values of $k$ and $\lambda$, respectively; see \s{simulations} and \s{hessian}.}
  \label{fig3} 
\end{figure}
\begin{figure*}
  \includegraphics[width=0.7\linewidth]{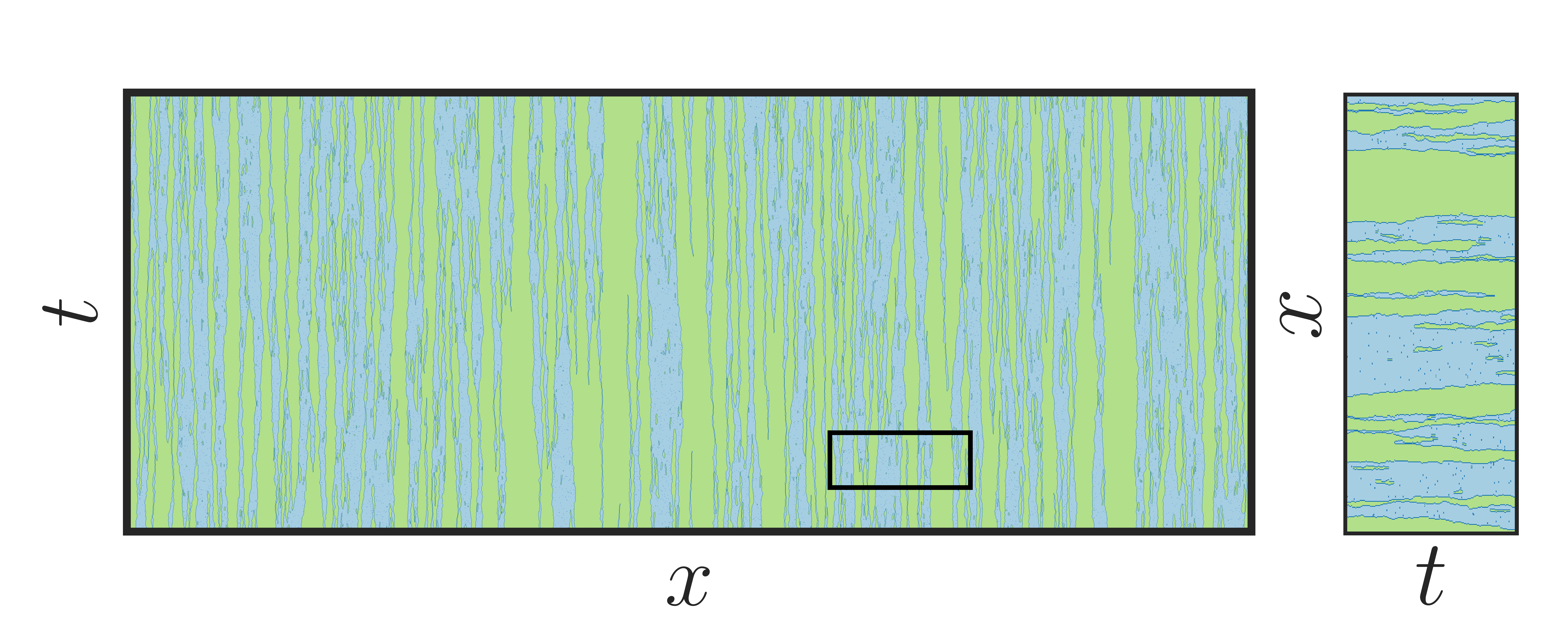}
  \caption{Space $x$ versus time $t$ plot of a trajectory of our statistical mechanical model of e-2--Mg$_2$(dobpdc), at reduced temperature $T^\star = 0.24$ (i.e. at the inflection point of the cooperative isotherm; see \f{fig3}(b)). Green indicates unoccupied sites or isolated \coo molecules; blue indicates polymerized \coo molecules. The slow dynamics associated with the diffusive fluctuations of chain lengths results in the large relaxation times shown in \f{fig3}(b) and the hysteresis shown in \f{fig2}(b). The right-hand panel is a zoom of the boxed region. See also \f{fig_ev}.}
  \label{fig4}
\end{figure*}

In \f{fig2}(a) we show experimental adsorption and desorption curves for \coo in the representative diamine-appended MOF e-2--Mg$_2$(dobpdc); experimental details are given in \s{experiments}. In the figure we also show early-stop desorption curves (dotted lines) obtained by reversing the temperature scan partway up the adsorption curve. These curves behave according to the scenario shown in panel (b) of \f{fig1}: the dynamics of cooperative adsorption in this MOF are qualitatively consistent with loading that proceeds in the absence of a phase transition. Similar non-monotonic behavior of the desorption curve is predicted to occur when diffusion of gas molecules within the framework is much slower than experimental timescales\c{Ravikovitch2005}; however, measured diffusion rates within our system\c{Forse2018} rule out this alternative scenario.

\subsection{Simulations provide a microscopic understanding of experimental observations}

In \f{fig2}(b) we show analogous data obtained from dynamic simulations of the statistical mechanical model of \cc{Kundu2018}. This model reproduces the cooperative adsorption of \coo within the diamine-appended MOFs by considering the polymerization of ammonium carbamate chains along the pore direction. Our prior analysis of the model indicates that an abrupt (but finite) increase of the mean length of polymerized \coo chains with temperature gives rise to a  step-like isotherm in the absence of a phase transition. The associated adsorption curves agree qualitatively and semi-quantitatively with experimental data\c{Kundu2018}. Here we present the model's dynamical behavior, and show that it too is consistent with experimental data: the simulations of \f{fig2}(b) are qualitatively consistent with the experiments of \f{fig2}(a), and with the mechanism summarized by \f{fig1}(b). Simulation results are averaged over 120 independent trajectories; error bars are smaller than the thickness of the lines. For computational efficiency we performed simulations at a higher temperature range ($540 - 750$ K) than experiments ($\approx 380-420$ K). The solid line marked ``equilibrium'' in \f{fig2}(b) is obtained by transfer-matrix calculation, while the colored lines are obtained by dynamical simulation. 

We show in \f{fig_rates} that varying the rates of temperature scan in experiment and simulation do not qualitatively change the scenario presented: in both cases, scanning temperature more slowly causes hysteresis loops to narrow, approaching (but not reaching) the equilibrium isobar. These changes are seen for any variation in scan rate, further supporting the picture we are presenting: modifying the hysteresis resulting from an underlying phase transition requires large changes in scan rate. (We also verified that simulations of the model reveal similar hysteretic behavior in uptake vs chemical potential curves at constant temperature; see \f{fig_mu}.).

Microscopic understanding of this phenomenon can be obtained by considering in more detail the model, sketched in \f{fig3}(a). The model is a representation of a one-dimensional channel of a diamine-appended MOF. The degrees of freedom of the model relate to the \coo binding states of the appended diamines, and we distinguish between \coo molecules bound singly or as part of a polymerized chain. The energetic parameters of the model are chosen to represent Mg metal sites, and the dynamics of the model is propagated using a standard continuous-time Monte Carlo scheme\c{Gibson2000}; details are given in \s{simulations}. 

The model reveals that hysteresis results from the ``flatness'' of the free-energy landscape near the inflection point of the isotherm. In \f{fig3}(b) we show the equilibrium isobar of the model (gray), together with a measure of the basic collective timescale of the system (red) and a measure of the curvature of the free-energy surface (blue). The timescale is determined by changing pressure abruptly and measuring the rate of the system's relaxation to equilibrium (\s{simulations}). The measure of free-energy curvature, $\lambda$, is the smallest eigenvalue of the Hessian matrix of the free energy expressed as a function of occupancies of \coo in various binding states (see Appendix~\ref{hessian}). Near the inflection point of the isobar the system possesses a free-energy landscape that is almost flat, with no strong thermodynamic driving force for \coo chains to grow or shrink. As a result, the relaxation time of the system is long, giving rise to hysteresis in the representation of \f{fig2}. 

The flatness of the free-energy landscape near the inflection point of the isotherm results in a slow diffusive dynamics of ammonium carbamate chain growth and shrinkage. In \f{fig4} we show a space-time plot of the model's dynamics, in which slow diffusive fluctuations of the lengths of polymerized \coo chains are apparent. By contrast, hysteresis in the presence of a first-order phase transition results from slow nucleation of the stable phase in a background of the metastable one.

\section{Conclusions} 

Cooperative adsorption of gases is of considerable scientific and technological importance. Despite recent advances in developing materials in which cooperative adsorption occurs\c{mcdonald2012,david2015,long_mixture}, we lack a complete understanding of how the phenomenon results from the interplay of gas molecules with their host framework. Building on \cc{Kundu2018}, this paper puts foward a molecular description of cooperative \coo adsorption in the metal--organic frameworks Mg$_2$(dobpdc), using the hysteresis seen in experiment as a means of distinguishing between two possible scenarios. Fundamental understanding of cooperative adsorbers will enable the rational design of these materials; for instance, analysis of the model allows us to identify the gas-metal binding energies for which cooperative adsorption occurs, and indicates how to move the inflection point of the isotherm to desired values of temperature and pressure. Fundamental understanding of gas uptake will also allow the design of experimental loading protocols that minimize dissipation\c{seifert2005entropy}, a first step toward the design of energy-efficient industrial protocols.

\acknowledgements

The computational portion of this research was carried out as part of a User project at the Molecular Foundry at Lawrence Berkeley National Laboratory (LBNL), supported by the Office of Science, Office of Basic Energy Sciences, of the U.S. Department of Energy under Contract No. DE-AC02-05CH11231. The experimental portions of the research and the contributions of J.E., R.L.S., Z.P., J.K., and J.R.L. were supported by the Center for Gas Separations, an Energy Frontier Research Center supported by the U.S. Department of Energy, Office of Science, Office of Basic Energy Sciences, under Award DE-SC0001015. J.R.L. serves as a director of and has a financial interest in Mosaic Materials, Inc., a start-up company working to commercialize metal-organic frameworks of the type investigated here. Some of these materials are the subject of patent applications submitted by the University of California, Berkeley. Data availability: All data needed to evaluate the conclusions in the paper are present in the paper and/or the Supplementary Materials. Additional data related to this paper may be requested from the authors. Author contributions: J.E., J.K., Z.P. and S.W. planned and carried out the simulations, and R.L.S. and J.R.L. did the same for the experiments. All authors contributed to the data analysis and interpretation.

\appendix
\widetext
\setlength{\parindent}{0em}
\setlength{\parskip}{1em}
\renewcommand\thefigure{A\arabic{figure}}    
\setcounter{figure}{0} 
\section{Experimental Methods}
\label{experiments}

\subsection{General Materials and Methods}
All synthetic manipulations were carried out under air. All solvents and the diamine e-2 ($N$-ethylethylenediamine) were purchased from commercial suppliers and used without further purification. The ligand 4,4'-dihydroxy-(1,1'-biphenyl)-3,3'-dicarboxylic acid (H$_4$dobpdc) was obtained from Hangzhou Trylead Chemical Technology Co. Ultra-high purity gases ($>$99.998\%) were used for all adsorption experiments. 
\subsection{Synthesis of e-2--Mg$_2$(dobpdc)}
The metal--organic framework Mg$_2$(dobpdc) was synthesized, washed, and characterized following a previously reported procedure\c{Siegelman2017}. Post-synthetic functionalization to prepare the diamine-appended framework e-2--Mg$_2$(dobpdc) was performed as reported previously\c{Siegelman2017}. The diamine loading was determined following literature procedure\c{Siegelman2017} by collecting $^{1}$H NMR spectra of material digested with DCl (35 wt \% in D$_2$O) in DMSO-{\em d}$_6$. Spectra were collected on a Bruker AMX 300 MHz NMR spectrometer and referenced to residual DMSO ($\delta$ 2.50 ppm). The diamine loading of as-synthesized e-2--Mg$_2$(dobpdc) was found to be 125\%, as determined from the ratio of the diamine to ligand peak integrals. A representative diamine loading of 98\% was determined following isobar collection. All adsorption data were collected on individual aliquots of a single sample within one week of preparation.
\subsection{Thermogravimetric Analysis}
Adsorption and desorption isobars were collected using a TGA Q5000 thermogravimetric analyzer. A flow rate of 10 mL/min was used for all gases, and masses were uncorrected for buoyancy effects. Samples were activated at 120$^\circ$C for 20 min under pure N$_2$ before isobar collection. Isobars were measured under pure CO$_2$ at atmospheric pressure using a temperature ramp rate of 1$^\circ$C/min.

\begin{table}[b]
   \begin{tabular}{c|c}
     Event & Rate \\
     \hline
     \et{U}{S} & $\omega_0 \exp [\beta (\mu - E_{\rm bs})]$ \\
     \et{S}{U} & $\omega_0 \exp [\beta (E_{\rm S} - E_{\rm bs})]$ \\
     \hline
     \et{SS[I]}{EE[A]} & $\omega_0 \exp [ -\beta E_{\rm bb}]$ \\
     \et{EE[A]}{SS[I]} & $\omega_0 \exp [  \beta (2(E_{\rm E}-E_{\rm S})-E_{\rm bb})]$ \\
     \et{EE[I]}{MM[A]} & $\omega_0 \exp [ -\beta E_{\rm bb}]$ \\
     \et{MM[A]}{EE[I]} & $\omega_0 \exp [  \beta (2(E_{\rm M}-E_{\rm E})-E_{\rm bb})]$ \\
     \et{SE[I]}{EM[A]} & $\omega_0 \exp [ -\beta E_{\rm bb}]$ \\
     \et{EM[A]}{SE[I]} & $\omega_0 \exp [  \beta (E_{\rm M}-E_{\rm S}-E_{\rm bb})]$ \\
     \et{ES[I]}{ME[A]} & $\omega_0 \exp [ -\beta E_{\rm bb}]$ \\
     \et{ME[A]}{ES[I]} & $\omega_0 \exp [  \beta (E_{\rm M}-E_{\rm S}-E_{\rm bb})]$ \\
       \end{tabular}
   \caption{\label{rate_table} Model rates: see \f{fig3}(a). We allow processes involving single sites (first two lines) or pairs of neighboring sites and their adjoining bond (subsequent lines). The energetic parameters are $E_{\rm S} = -1.0 \epsilon$, $E_{\rm E} = -2.592\epsilon$ , $E_{\rm M}=-3.24\epsilon$, $\beta \mu =-13.5$, $E_{\rm bs} = 0.5 \epsilon$, and $E_{\rm bb} = \epsilon$. Here $\beta = \epsilon/(k_{\rm B}T)$. The parameter $\epsilon=22.6$ kJ/mol is binding energy of a single \coo molecule in the MOF, obtained using quantum mechanical density-functional theory\c{Kundu2018}.}
\end{table}

\section{Simulations of the statistical mechanical model}
\label{simulations}
We used the one-dimension version of the model of \cc{Kundu2018}, sketched in \f{fig3}(a), choosing energetic parameters appropriate for the metal Mg. We simulated the model using the Next Reaction Method of Gibson and Bruck\c{Gibson2000,Li2015}, whose computational cost per event scales as $\ln N$ for a lattice model of $N$ sites. We allowed the range of processes described in Table~\ref{rate_table}, including the binding and unbinding of \coo molecules, and the formation of bonds between them. The stationary distribution of the Markov chain defined by these rates is the grand-canonical probability density of the system. 

The constants $\omega_0$ and $\epsilon$ set the time- and energy scales of the system. The energy scale $\epsilon=22.6$ kJ/mol is the binding energy of a single \coo molecule within the framework. Following \cc{Kundu2018}, the binding constants of the gas molecules in the occupied S, M, E states (see \f{fig3}(a)) are $E_{\rm S} = -1.0 \epsilon$, $E_{\rm E} = -2.592\epsilon$ and $E_{\rm M}=-3.24\epsilon$. We set the activity of the system to be $ \beta \mu = -13.5$. We set the barrier height for the site transition to $E_{\rm bs} = 0.5 \epsilon$ and the barrier height for all bond transitions to $E_{\rm bb} = \epsilon$, and verified via numerical simulations that small variations in these choices do not affect any of the conclusions of our work. For reasons of computational feasibility we performed the simulations at a higher temperature range than the experiments. 

We measured the relaxation rate $k$ in our simulations, displayed in Fig. 3, as follows. The system is initially equilibrated at a certain value of temperature $T_0$ and activity or pressure $P_0$, and has a \coo occupancy $\rho_0$. We then make an abrupt change in temperature of the of the system. Following this change we measured the \coo occupancy $\rho(t)$, averaged over 120 trajectories, until equilibrium is achieved. We compared the final density $\rho_f$ measured in simulations with exact analytical transfer-matrix calculations to ensure that the system has equilibrated. We fit $\bar{\rho}(t) = (\rho_f - \rho(t))/(\rho_f - \rho_0)$ to an exponential form ${\rm e}^{-k t}$, thus obtaining $k$. Values obtained in this way are shown in Fig. 3. We determined the basic time scale of the system, $\omega_0 = 210$ s$^{-1}$, by equating the relaxation rate measured in experiments (upon an abrupt pressure change) to the relaxation rate of our model $k$ for a step change in pressure at constant temperature.

\section{Analytic treatment of the statistical mechanical model}
\label{hessian}

 As shown in \f{fig3}, the curvature of the free-energy landscape of our statistical mechanical model is correlated with the dynamical relaxation time of the model. Near the inflection point of the isobar the free-energy landscape is almost flat, implying that the thermodynamic driving force for polymerized \coo chains to grow or shrink is small. In this section we detail our calculation of the curvature of the model's free energy.

 The grand partition function of the model is given by
 \begin{equation}
 	\mathcal{Z}=\sum_{\{n_1,n_{\rm int},n_{\rm end}\}} K_1^{n_1} K_{\rm end}^{n_{\rm end}} K_{\rm int}^{n_{\rm int}} \Gamma (n_1,n_{\rm int},n_{\rm end}),
 \end{equation}
 where

\begin{eqnarray}
 	\Gamma(n_1,n_{\rm int},n_{\rm end})=
	 \frac{(N-n_{\rm int}-n_{\rm end}/2)!}{(N-n_1-n_{\rm int}-n_{\rm end})!(n_{\rm end}/2)!n_1!} 
 	\cdot \frac{(n_{\rm end}/2+n_{\rm int}-1)!}{(n_{\rm end}/2-1)! n_{\rm int}!}
 	\label{eq_gamma}
\end{eqnarray}

is the number of ways of arranging $n_1$ single molecules, $n_{\rm int}$ internal chain molecules, and $n_{\rm end}$ chain end-points on a one-dimensional lattice of $N$ sites. The quantity $K_{\alpha}$, with $\alpha \in \{1, {\rm int}, {\rm end}\}$, is the statistical weight of a \coo molecule in conformation $\alpha$. The free energy in the thermodynamic limit is

\begin{eqnarray}
 	f(x_1,x_{\rm int},x_{\rm end})&=&-(1-x_{\rm int} -\frac{x_{\rm end}}{2}) \ln (1-x_{\rm int} 
	-\frac{x_{\rm end}}{2}) \nonumber \\
 	&-&(x_{\rm int} +\frac{x_{\rm end}}{2}) \ln (x_{\rm int}+\frac{x_{\rm end}}{2})+x_1 \ln x_1+x_{\rm int} \ln x_{\rm int}+x_{\rm end} \ln \frac{x_{\rm end}}{2} \nonumber\\
 	&+&(1-x_1-x_{\rm int}-x_{\rm end}) \ln (1-x_1-x_{\rm int}-x_{\rm end}),
\end{eqnarray}

where $x_{\alpha}$, with $\alpha \in \{1, {\rm int}, {\rm end}\}$, is the fraction of \coo molecules in conformation $\alpha$. 

The Hessian $H$ is a $3\times 3$ matrix built from the three variables $x_1$, $x_{\rm int}$, and $x_{\rm end}$, with matrix elements 
 \begin{equation}
H_{\alpha\beta}=\frac{\delta^2 f(x_1,x_{\rm int},x_{\rm end})}{\delta x_{\alpha} \delta x_{\beta}}{\Bigg |}_{\rm min},
 \end{equation}
 evaluated at the free-energy minimum. 

For a given pressure (the statistical weight $K_\alpha$ is proportional to the pressure), we evaluated $H_{\alpha \beta}$ at the equilibrium values of $x_1$, $x_{\rm int}$, and $x_{\rm end}$, and calculated the three eigenvalues. We find that the smallest eigenvalue, called $\lambda$ in \f{fig3}(b), is small at the inflection point of the isobar, indicating that there is no strong thermodynamic driving force for the growth or decay of ammonium carbamate chains. As a result, the dynamics of the model becomes slow, and we show in \f{fig3}(b) that the minimum relaxation rate $k$ occurs near the minimum of $\lambda$.

\section{Additional figures}

\begin{figure*}
  \begin{center}
    \includegraphics[width=0.8\textwidth,keepaspectratio]{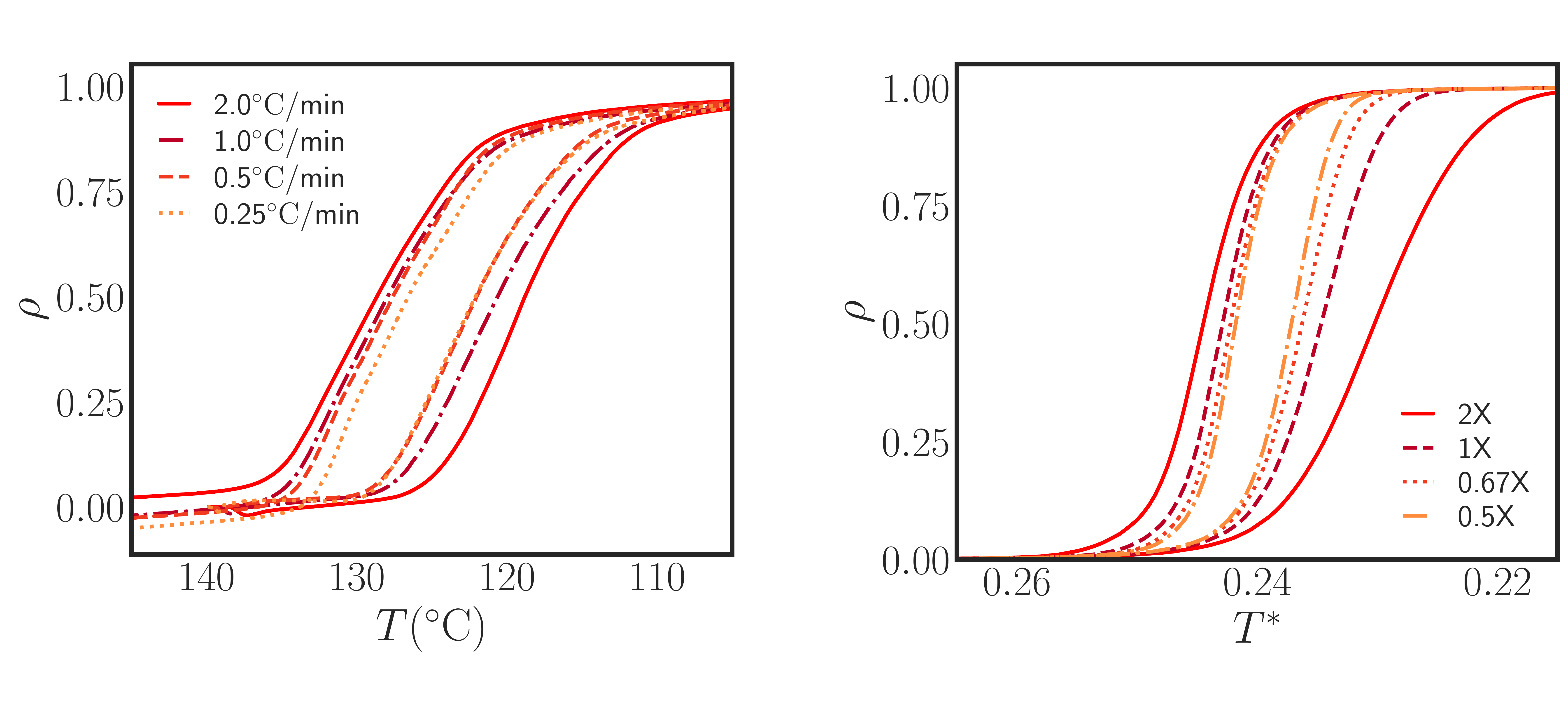}
  \end{center}
  \vspace{-20pt}
  \caption{As \f{fig2}(b), for different rates of temperature scan. Experiment and simulation are qualitatively consistent.}
  \label{fig_rates}
\end{figure*}

\begin{figure}
  \begin{center}
    \includegraphics[width=0.49\textwidth,keepaspectratio]{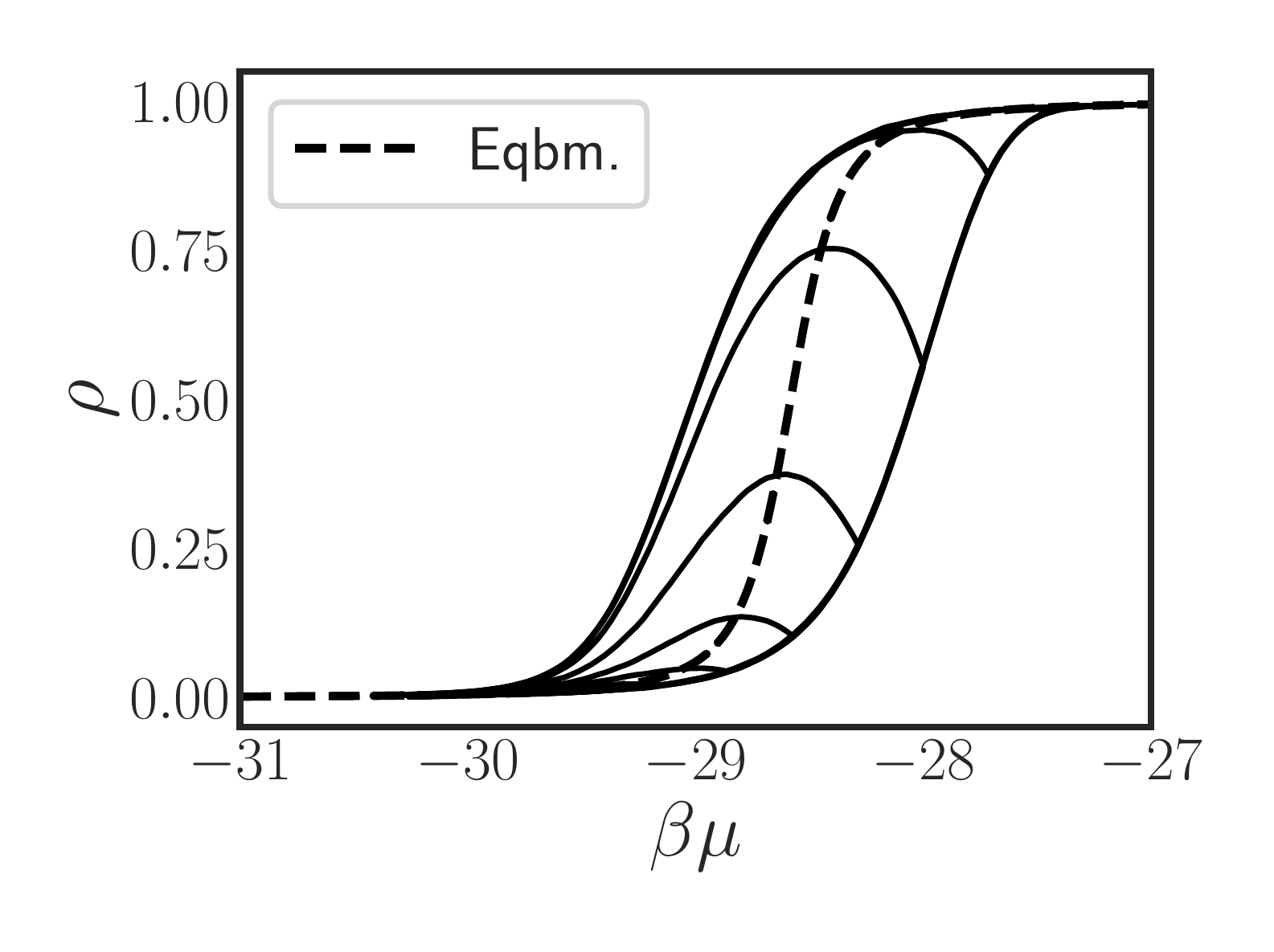}
  \end{center}
  \vspace{-20pt}
  \caption{As \f{fig2}(b), but as a function of chemical potential  $\beta \mu$ instead of temperature; the behavior is qualitatively similar.  Model parameters: $E_s = 1.0$, $E_m = 2.5$, $E_t = 2.0$ and $T^* = 0.175$.}
  \label{fig_mu}
\end{figure}

\begin{figure*}
  \begin{center}
    \includegraphics[width=0.95\textwidth]{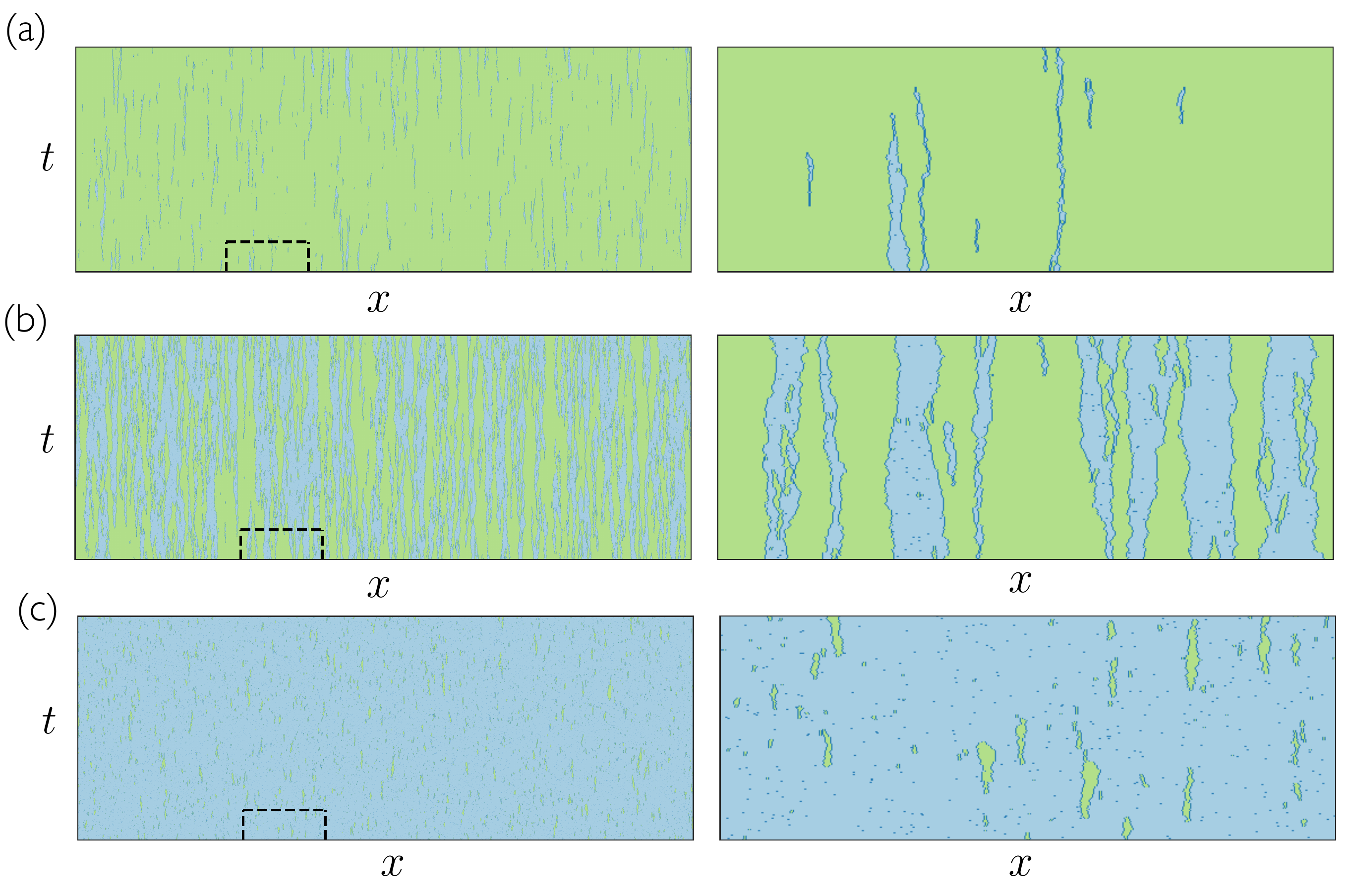}
  \end{center}
  \caption{Space-time plots of the model under a sudden change of chemical potential to the values (a) $\beta \mu = 29.14$, (b) $\beta \mu = -28.57$, and (c)  $\beta \mu = -28.0$, which lie at and on either side of the inflection point of the curve shown in \f{fig_mu}. Green represents unoccupied sites or isolated \coo molecules; blue represents polymerized \coo molecules. The right-hand panels are englargements of the boxed regions.}
  \label{fig_ev}
\end{figure*}

\begin{itemize}
\item In \f{fig_rates} we show experiments and simulations similar to \f{fig2}, but for different rates of temperate scan.

\item In \f{fig_mu} we show uptake vs chemical potential at constant temperature. At each value of chemical potential the system evolves for an observation time of $10^8$ before the chemical potential is changed. The dashed line gives the exact equilibrium density and is computed using transfer matrices. The behavior of the scanning curves is qualitatively similar to the case of uptake versus temperature at fixed pressure. 

\item In \f{fig_ev} we show space-time plots of the model at three different points on the curve shown in \f{fig_mu}.
\end{itemize}


\end{document}